# SecureReg: Combining NLP and MLP for Enhanced Detection of Malicious Domain Name Registrations

Furkan Çolhak
CCIP, Center for Cyber Security and
Critical Infrastructure Protection,
Kadir Has University
Istanbul, Turkey
furkancolhak@stu.khas.edu.tr

Mert İlhan Ecevit
CCIP, Center for Cyber Security and
Critical Infrastructure Protection,
Kadir Has University
Istanbul, Turkey
mertilhan.ecevit@khas.edu.tr

Hasan Dağ
CCIP, Center for Cyber Security and
Critical Infrastructure Protection,
Kadir Has University
Istanbul, Turkey
hasan.dag@khas.edu.tr

Reiner Creutzburg
SRH Berlin University of Applied Technology, Berlin School of Technology, Berlin, Germany
Technische Hochschule Brandenburg, Fachbereich Informatik und Medien, Brandenburg, Germany
reiner.creutzburg@srh.de & creutzburg@th-brandenburg.de

*Abstract*—The escalating landscape of cyber threats, characterized by the registration of thousands of new domains daily for large-scale Internet attacks such as spam, phishing, and drive-by downloads, underscores the imperative for innovative detection methodologies. This paper introduces a cutting-edge approach for identifying suspicious domains at the onset of the registration process. The accompanying data pipeline generates crucial features by comparing new domains to registered domains, emphasizing the crucial similarity score. The proposed system analyzes semantic and numerical attributes by leveraging a novel combination of Natural Language Processing (NLP) techniques, including a pretrained CANINE model and Multilayer Perceptron (MLP) models, providing a robust solution for early threat detection. This integrated Pretrained NLP (CANINE) + MLP model showcases the outstanding performance, surpassing both individual pretrained NLP models and standalone MLP models. With an F1 score of 84.86% and an accuracy of 84.95% on the SecureReg dataset, it effectively detects malicious domain registrations. The findings demonstrate the effectiveness of the integrated approach and contribute to the ongoing efforts to develop proactive strategies to mitigate the risks associated with illicit online activities through the early identification of suspicious domain registrations.

## I. INTRODUCTION

The Domain Name System (DNS) functions as the foundational infrastructure of the Internet, comparable to an extensive directory facilitating the translation of website names into numerical IP addresses understandable to computers. This system is fundamental in enabling seamless online communication and access to web resources. However, despite its critical importance, DNS is constantly threatened by cybercriminals who seek to exploit its vulnerabilities for vile purposes.

In recent years, there has been a concerning rise in attacks targeting the DNS infrastructure. Malicious actors leverage various techniques, including DNS spoofing, cache poisoning, and distributed denial-of-service (DDoS) attacks [1], to disrupt the normal functioning of DNS and redirect users to fraudulent websites. These vicious activities pose significant risks to Internet users, ranging from financial losses due to scams to compromising sensitive data through malware infections. Recent statistics from the Domain Abuse Activity Reporting (DAAR) [2] system underscore the pervasive nature of DNS abuse, with a significant number of generic Top-Level Domains (gTLDs) harboring security threats.

In response to these challenges, domain registries have developed various solutions. EURid's Premadoma [3] employs machine learning to focus on large-scale malicious campaigns, while DNS Belgium and SWITCH use scoring systems based on static rules. Nominet's Domain Watch [4] specifically targets phishing scams. RegCheck inspired this study [5], in which the data-driven base model employs logistic regression, a machine learning algorithm, to assign weights to the risk factors using labeled registrations. Despite the effectiveness of these systems, there is a shared recognition that a more nuanced and customized approach is needed.

In response to this need, SecureReg has been designed as a novel system to assign interpretable risk scores to new domain name registrations. Unlike existing methods, SecureReg incorporates a unique approach beyond traditional risk assessment. It utilizes feature extraction, Natural Language Processing (NLP), and Multilayer Perceptron (MLP) models to assess the risk associated with a new domain and determine its similarity score to the registrant's database at registration.

The conducted research has contributed in the following ways to this study;

- A proactive time-of-registration domain reputation system has been developed, which combines common features and similarity scores to identify potentially malicious domain registrations at the point of registration. This approach enhances the ability to detect threats early on.

- an advanced model design that utilizes Pretrained NLP and MLP model has been created, strengthening the effectiveness of the proactive domain reputation system in identifying potential malicious domains."

The remaining sections of this study are organized as follows: Section II reviews related work. The proposed approach is presented in Section III. Section IV details the data pipeline, including dataset description and the extraction of features. Section V covers the model pipeline and structure. Section VI analyzes results, including evaluation metrics and performance. Section VII discusses limitations. Finally, Section VIII concludes, summarizing findings and suggesting future work.

## II. RELATED WORK

This section overviews various approaches categorized into existing solutions that aim to detect potentially malicious domain registrations during the registration process, such as NLP-based Studies, MLP-based Studies, and Combined MLP+NLP Studies. It highlights vital studies and methodologies within each category, emphasizing the evolving landscape of identifying potentially malicious domain registrations at the point of registration.

**Existing Solutions:** In this section, the current approaches employed by European registries have been overviewed. EURid (.eu) employs the Premadoma system, thoroughly evaluating and leveraging machine learning to target large-scale malicious campaigns [3]. However, Premadoma primarily focuses on such campaigns, necessitating additional systems to detect scammers registering single suspect domain names. DNS Belgium (.be) and SWITCH (.ch, .li) utilize a scoring system based on static rules to identify potentially malicious registrations, though the static nature limits customization. DNS Belgium is exploring the integration of machine learning to enhance adaptability, and ongoing collaboration between the registries is maintained. Nominet (.uk) introduces Domain Watch, targeting potential phishing scams and suspending potentially malicious domains shortly after registration [4]. However, due to Nominet's cautious approach, limited information is available on Domain Watch's inner workings and performance. Regcheck offers a transparent design, enabling other registries to adopt a similar approach [5]. Like Nominet's cautious approach with Domain Watch, Regcheck keeps risk factors private to prevent exploitation by adversaries. Regcheck allows registries to develop and evaluate statistical models assigning risk scores to domain name registrations, indicating the likelihood of hosting abusive websites.

**NLP-based Studies:** Several studies have investigated methods for detecting malicious domains using Natural Language Processing (NLP) techniques combined with machine learning or deep learning. Pradeepa and Devi (2022) offer a comprehensive review, emphasizing the importance of accurately identifying malicious domain names amidst evolving attacker tactics [6]. Gogoi and Ahmed (2023) propose a transformer-based approach to detect Domain Generation Algorithm (DGA)-generated domain names with high accuracy [7]. Chang, Du, and Wang (2021) introduce a robust method using BERT for malicious URL detection, achieving high accuracy, recall rate, and F1 value [8]. Yang et al. (2022) propose N-Trans, integrating N-gram and Transformer models, demonstrating superior performance in distinguishing between legitimate and malicious domain names [9].

**MLP-based Studies:** Multilayer Perceptron (MLP) models have been extensively explored to identify potentially malicious domains and URLs. Sharma (2020) proposes a method for detecting spam domains using machine learning, leveraging characteristics such as DKIM signature domain and active DNS records like SPF records and Authoritative nameservers [10]. Vranken and Alizadeh (2022) address the detection of domain names generated by domain name generation algorithms (DGAs) using machine learning and deep learning techniques, achieving high performance in distinguishing DGA-generated domain names [11]. SHOID (2018) focuses on classifying malicious URLs using the Multilayer Perceptron technique, aiming to enhance web application security and provide users with safer browsing experiences [12].

**Combined MLP+NLP Studies:** Recent advancements in cybersecurity research have integrated NLP and MLP techniques to enhance the detection of malicious activities. Jishnu and Arthi (2023) propose using BERT and URL feature extraction to detect phishing URLs, achieving 97.32% accuracy [13]. Düzgün et al. (2023) address network intrusion detection by combining tabular and text-based features with pre-trained transformer models, outperforming traditional methods [14]. Çolhak et al. (2024) focus on phishing website detection, using a specialized MLP model and two pretrained NLP models to analyze HTML content. Their approach achieves impressive results by fusing the embeddings from these models and inputting them into a linear classifier, outperforming existing methods on their MTLP and the CatchPhish HTML datasets. [15].

When the existing solutions used during the registration process have been examined, it becomes evident that there's a common understanding that a more detailed and tailored approach is necessary. However, despite the effectiveness of these systems, there's still a noticeable gap, especially in adopting up-to-date models. This work addresses this gap by presenting a novel solution for detecting malicious domain names at the time of registration, enhancing performance through the combination of Pretrained NLP and MLP models.

## III. APPROACH DESIGN

The process begins with registering a new domain, such as "l1nkedin.(TLD)". This newly registered domain enters the data pipeline, where it undergoes feature extraction and similarity score calculation by comparing it with the registrant database. Subsequently, these features are added to the primary dataset as a new column. The enriched data flows into a model pipeline that utilizes pre-trained NLP and MLP models. This evolved model generates an output classifying the domain as benign or suspicious. This final determination is communicated to the registry operator, who takes appropriate actions based on the categorization, ensuring the security and integrity of the domain registration system.

As illustrated in Fig. 1, the comprehensive approach is outlined herein.

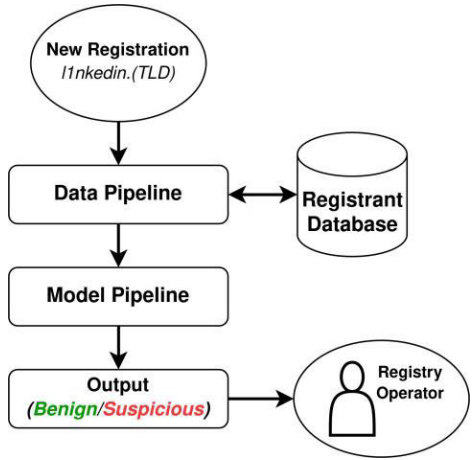

**Fig. 1.** Approach Design

## IV. DATA PIPELINE

This study developed a data pipeline to analyze a real-life database comprising benign and malicious domain names (without TLDs). Alexa's top 100k [16] served as the fictional registrant dataset. The pipeline enriched the primary dataset with standard features and similar features. This systematic approach enhances understanding of domain relationships, identifying potential threats, and facilitating data enrichment. The pipeline ensures efficient processing and offers valuable insights into domain similarities within the datasets.

Fig. 2 provides a brief overview of the data pipeline.

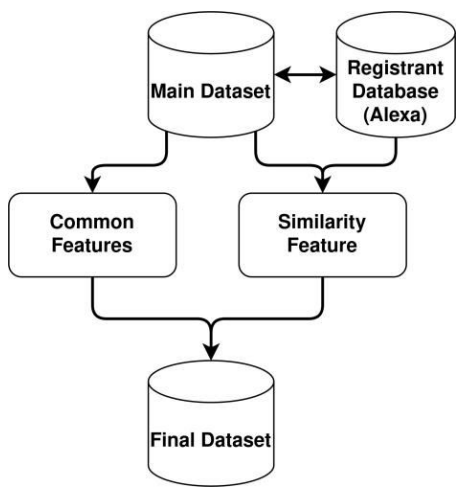

**Fig. 2.** Data Pipeline

### A. Dataset

Data has been gathered from two different sources to create the dataset. Initially, by randomly selecting benign (safe) website URLs from the Alexa 1m [16] dataset have been obtained. Then, malicious URLs from OpenPhish are gathered. Using a WHOIS tool designed explicitly for open-source intelligence (OSINT), domain names from these URLs and their top-level domains (TLDs) have been removed. Subsequently, duplicate domain names and those already present in the top 100k list on Alexa have been eliminated. This step was crucial as utilization of the Alexa top 100k [16] list as a fictional database of registrants took place. The dataset, named the SecReg Dataset, is available for access on the GitHub Repository [17].

Table I shows data distribution in the SecReg Dataset.

**TABLE I.** Data Distribution in SecReg Dataset

|  | Benign | Malicious | Total |
|---|---|---|---|
| **SecReg Dataset** | 4,198 | 4,702 | 8,900 |

### B. Feature Extraction

*1) Common Features:* This study leverages common features widely adopted in domain name and URL analysis research. These features, which have been utilized in numerous studies [18] [19] [20] within the field, serve as foundational elements for understanding and categorizing domain names effectively.

**Length:** Suspicious actors often create domain names that are excessively long or convoluted to obfuscate their true intentions. For instance, legitimate domains tend to be concise and directly relevant to their content, such as "examplebank.(TLD)" However, suspicious domains may exhibit unusual length, incorporating additional terms or hyphens to create confusion, as seen in examples like "securelogin-examplebank-accountverification.(TLD)" Analyzing the length of a domain name can help identify such anomalies and flag them for further scrutiny.

**Digit Count:** Many phishing or scam websites utilize numerical digits within their domain names to mimic legitimate domain names or to create a sense of urgency (e.g., "accountupdate123.(TLD)"). Monitoring the occurrence of numerical digits can aid in detecting these deceptive practices.

**Special Character Count:** Special Character Count: Suspicious domains may employ special characters in uncommon ways to evade detection or to impersonate legitimate websites (e.g., "b@nkofamerica.(TLD)". Tracking the presence of special characters can help identify such attempts at domain spoofing or phishing.

*2) Similarity Feature:* In this subsection, the Sequence Matcher method has been utilized to determine the similarity between newly registered domain names and entries in a registrant database. The Sequence Matcher is a Python module that compares sequences by identifying the longest contiguous matching subsequence between them [21], producing a similarity score.

- **Example:**
  - **String 1:** "example"
  - **String 2:** "ample"

The Sequence Matcher compares these two strings and identifies that "ample" is the longest contiguous matching subsequence between them. It then calculates a similarity score based on the length of this matching subsequence relative to the lengths of the input strings.

- **Calculation:**
  - Length of matching subsequence: 5
  - Total length of input strings: 7 (since "example" has 7 characters and "ample" has 5 characters)
  - Similarity score: 5/7 ≈ 0.71

This similarity score indicates that the strings "example" and "ample" are about 71% similar based on the longest contiguous matching subsequence.

In this study's specific use case, the Sequence Matcher method would be applied to compare a newly registered domain name with each entry in the registrant database. The domain with the highest similarity score would be considered the most similar to the newly registered domain and would be added to the primary dataset for further analysis.

## V. MODEL PIPELINE

The proposed model pipeline is a sophisticated multi-modal system for data classification, particularly in analyzing domain names. This choice is driven by the proven efficacy of pretrained models, which excel in diverse fields and consistently deliver high performance [22]. Additionally, the integration of Natural Language Processing (NLP) and Multilayer Perceptron (MLP) models in a multi-modal framework is selected for its capacity to provide a double-layer classification approach, enhancing the overall precision and interpretability of the system. The input data encompasses a primary textual feature, "domain_name," and several numeric features such as "similarity_score," "length," "digit_count," and "special_character_count." To harness the rich semantic information within the textual data, a Natural Language Processing (NLP) model is employed, while a Multilayer Perceptron (MLP) model processes the numeric features. Fig. 3 provides a clear visual representation of the entire process in the applied model pipeline, illustrating each step and how they work together.

### A. Natural Language Processing (NLP)

In the cybersecurity domain, Transformer models have demonstrated significant efficacy in various tasks, including malware detection [23], phishing detection [15], threat intelligence analysis [24], and log analysis [25]. Their ability to process and understand textual data at scale makes them invaluable for identifying malicious patterns, detecting anomalies, and enhancing overall security posture.

In this study, various Transformer models have been applied, including:

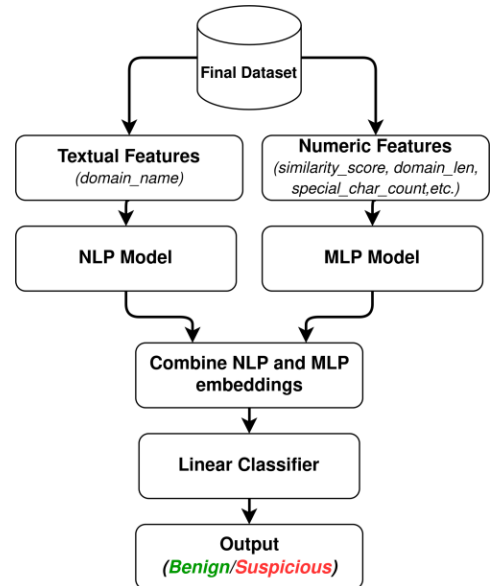

**Fig. 3.** Model Pipeline

**ALBERT:** ALBERT addresses the computational inefficiencies of BERT by introducing parameter reduction techniques such as factorized embedding parameterization and cross-layer parameter sharing, resulting in a more efficient yet powerful model [26].

**BERT**: BERT revolutionized NLP by introducing bidirectional context encoding, capturing deeper semantic understanding by considering both left and proper context during pretraining [27].

**CANINE:** CANINE adopts a character-level encoding approach, bypassing traditional tokenization, which can be particularly beneficial for handling languages with complex morphology and tokenization challenges [28].

**RoBERTa:** RoBERTa builds upon BERT's architecture by optimizing pretraining objectives and training strategies, enhancing performance across various NLP tasks [29].

**MobileBERT:** MobileBERT, a compact variant of BERT, optimizes for mobile and edge devices by employing techniques like knowledge distillation and network pruning, ensuring high performance with reduced computational cost [30].

Throughout this study, these Transformer models are utilized to evaluate their effectiveness in processing and extracting insights from textual data, explicitly focusing on the "domain_name" column within SecReg dataset [17].

### B. Multilayer Perceptron (MLP)

In the MLP part, it is acknowledged that the extracted numeric features are not highly complicated. Therefore, a requirement for the design of a very basic Multilayer Perceptron (MLP) structure has emerged. The simplicity of the features allows for a straightforward model architecture—an artificial neural network with multiple layers of nodes that processes input information [31] and learns patterns through weighted connections and activation functions, contributing to the overall model's ability to detect and prevent malicious domain registrations. The simplicity of the features allows for a straightforward model architecture, as outlined in Table II.

**TABLE II.** MLP Structure

| Layer | Dimensions | Activation | Regularization |
|---|---|---|---|
| Input | [*input_dim*] | | |
| FC1 | [*input_dim*, 1024] | LeakyReLU (0.1) | BN |
| Dropout | $p = 0.2$ | | |
| FC2 | [1024, 2056] | LeakyReLU (0.1) | BN |
| Dropout | $p = 0.2$ | | |
| FC3 | [2056, 512] | LeakyReLU (0.1) | BN |
| Dropout | $p = 0.2$ | | |
| FC4 | [512, 16] | LeakyReLU (0.1) | BN |
| Dropout | $p = 0.3$ | | |
| FC5 | [16, *output_dim*] | | |

**Note:** [*input_dim*] corresponds to the number of numeric columns in the input data, and *output_dim* = 2 represents benign/suspicious classification.

Finally, these models' outputs are combined through an embedding layer, facilitating the integration of information derived from textual and numeric domains. The final stage involves a linear classifier, which leverages the fused representation to make a binary classification decision, ultimately categorizing the input data as "benign" or "suspicious".

## VI. RESULTS

### A. Evaluation Metrics

Evaluating detection methods for benign and malicious domain names is essential for assessing their effectiveness. Table III presents various metrics formulas and their descriptions, which are instrumental in this evaluation process.

**TABLE III.** Metrics Formulas and Descriptions for Detecting Benign and Malicious Domain Names

| Metric | Formula | Description |
|---|---|---|
| Accuracy | $\frac{TP+TN}{TP+TN+FP+FN} \times 100$ | Percentage of correctly classified benign and malicious domain names out of the total instances. |
| Precision | $\frac{TP}{TP+FP}$ | Proportion of correctly predicted malicious domain names among those predicted as malicious. |
| Recall | $\frac{TP}{TP+FN}$ | Proportion of correctly predicted malicious domain names out of all actual malicious domain names. |
| F1-Score | $2 \times \frac{Precision \times Recall}{Precision+Recall}$ | Harmonic mean of precision and recall, providing a balanced measure that considers both false positives and false negatives. |

### B. Experimental Setup

The experiments used an Intel Core i9-10900x processor running at 3.70 GHz and an NVIDIA RTX A4000 GPU with 16GB of ECC memory.

Additionally, the dataset is partitioned into training (70%), validation (15%), and test (15%) sets to evaluate the models' performance thoroughly. All performance comparisons for each model were conducted under the same parameters.

### C. Performance Evaluation

This section presents the performance evaluation of various models for identifying suspicious domains during registration, including pretrained Natural Language Processing (NLP) models and Multilayer Perceptron (MLP) models.

**TABLE IV.** Performance Comparison of Pretrained Models of Accuracy (Acc), F1-Scores (F1), Precision (Prec), and Recall (Rec)

| Pretrained Model | F1 (%) | Acc (%) | Prec (%) | Rec (%) |
|---|---|---|---|---|
| ALBERT | 81.06 | 80.22 | 81.48 | 80.82 |
| BERT | 84.17 | 83.45 | 83.73 | 84.93 |
| **CANINE** | **84.76** | **84.72** | **87.34** | **82.57** |
| MobileBERT | 82.76 | 81.95 | 83.00 | 82.89 |
| RoBERTa | 83.22 | 82.92 | 84.96 | 81.57 |

In Table IV, CANINE outperforms other pretrained NLP models regarding F1 score, accuracy, precision, and recall. This suggests that CANINE is particularly effective for identifying suspicious domains during registration.

**TABLE V.** Comparison between models of Accuracy (Acc), F1-Scores (F1), Precision (Prec), and Recall (Rec)

| Model | F1 (%) | Acc (%) | Prec (%) | Rec (%) |
|---|---|---|---|---|
| MLP | 71.26 | 65.02 | 62.66 | 83.25 |
| NLP (CANINE) | 84.76 | 84.72 | 87.34 | 82.57 |
| **NLP + MLP** | **84.86** | **84.95** | **88.81** | **81.56** |

In Table V, the combined NLP + MLP approach performs better than individual models, indicating that integrating NLP techniques with MLP models enhances the system's ability to detect suspicious domains early in the registration process. This integrated approach achieves higher precision and accuracy, crucial for reducing false positives and improving overall detection effectiveness.

## VII. LIMITATIONS

This approach faces several challenges. While existing methods typically employ light models, The implemented model is notably more sophisticated and large-scale. Consequently, computational demands and processing delays may hinder its application in real-time scenarios, such as during the domain registration process.

Another significant issue is the lack of publicly shared datasets for comparison. Because existing solutions haven't made their benchmark datasets available to the public, Comparing how well the applied method performs against theirs is impossible. This makes it challenging to determine the effectiveness of the designed approach compared to existing solutions.

## VIII. CONCLUSION AND FUTURE WORK

Malicious domains pose an increasingly significant daily threat, underscoring the critical need for proactive measures to enhance security. Detecting these domains during registration can be pivotal in fortifying defenses against cyber threats.

In light of this pressing concern, this research introduces SecureReg, a novel approach that shows significant promise in combating malicious domain registrations. While many studies have explored domain names, there's a notable lack of proactive prevention systems. The findings of this study fill this crucial gap by offering a solution that stands out for its innovation and effectiveness. Moreover, unlike existing methods that keep their datasets private, the SecReg dataset is openly shared with the community. This transparency fosters collaboration and contributes to comprehensive DNS research, benefiting the field.

SecureReg leverages a unique blend of feature extraction, Natural Language Processing (NLP), and Multilayer Perceptron (MLP) models, allowing it to analyze textual and numerical attributes during domain registration. The utilized Pretrained NLP (CANINE) + MLP model demonstrates remarkable performance, surpassing individual pretrained NLP models and standalone MLP models. It achieves an F1 score of 84.86% and an accuracy of 84.95% on the SecReg dataset, effectively identifying malicious domain registrations.

Moving forward, it is planned to collaborate with real-life domain registration data providers to evaluate SecureReg in real-life conditions. Additionally, The aim is to optimize the model's efficiency and reduce its overall capacity to address computational challenges effectively.

## IX. Acknowledgments

This work was supported partially by the European Union in the framework of ERASMUS MUNDUS, Project CyberMACS (Project #101082683) (https://cybermacs.eu).